\documentclass[iop,apjl,twocolappendix]{emulateapj}
\usepackage{hyperref}

\shorttitle{The formation of shell galaxies}
\shortauthors{A.P. Cooper et al.}

\begin{document}

\title{The formation of shell galaxies similar to NGC 7600 \\ in the
  cold dark matter cosmogony}

\author{Andrew P. Cooper}
\email{acooper@mpa-garching.mpg.de}
\affil{Max Planck Institut f\"{u}r Astrophysik, Karl-Schwarzschild-Str. 1  
85741 Garching, Germany}

\author{David Mart\'{i}nez-Delgado}
\affil{Max Planck Institut f\"{u}r  Astronomie, K\"{o}nigstuhl 17
D-69117, Heidelberg, Germany}
\email{delgado@mpia-hd.mpg.de}

\author{John Helly, Carlos Frenk, Shaun Cole} 
\affil{Institute For
    Computational Cosmology, Department of Physics, University of Durham,
    South Road, \mbox{DH1 3LE}, Durham, UK}

\author{Ken Crawford}
\affil{Rancho del Sol Observatory, Camino, CA 95709, USA}

\author{Stefano Zibetti}
\affil{Dark Cosmology Centre, Niels Bohr Institute - University of Copenhagen
Juliane Maries Vej 30, DK-2100 Copenhagen, Denmark}

\author{Julio A. Carballo-Bello}
\affil{Instituto de Astrof\'\i sica de Canarias, La Laguna, Spain}

\and 
\author{R. Jay Gabany}
\affil{Black Bird Observatory II, California, USA}

\begin{abstract}

We present new deep observations of `shell' structures in the halo
of the nearby elliptical galaxy NGC~7600, alongside a movie of galaxy
formation in a cold dark matter universe
(\url{http://www.virgo.dur.ac.uk/shell-galaxies}). The movie, based on
an ab initio cosmological simulation, shows how continuous accretion
of clumps of dark matter and stars creates a swath of diffuse
circumgalactic structures. The disruption of a massive clump on a
near-radial orbit creates a complex system of transient concentric
shells which bare a striking resemblance to those of NGC~7600. With
the aid of the simulation we interpret NGC~7600 in the context of the
CDM model.

\end{abstract}



\section{Introduction}

\begin{figure}
\center
\includegraphics[clip=True,width=8.0cm]{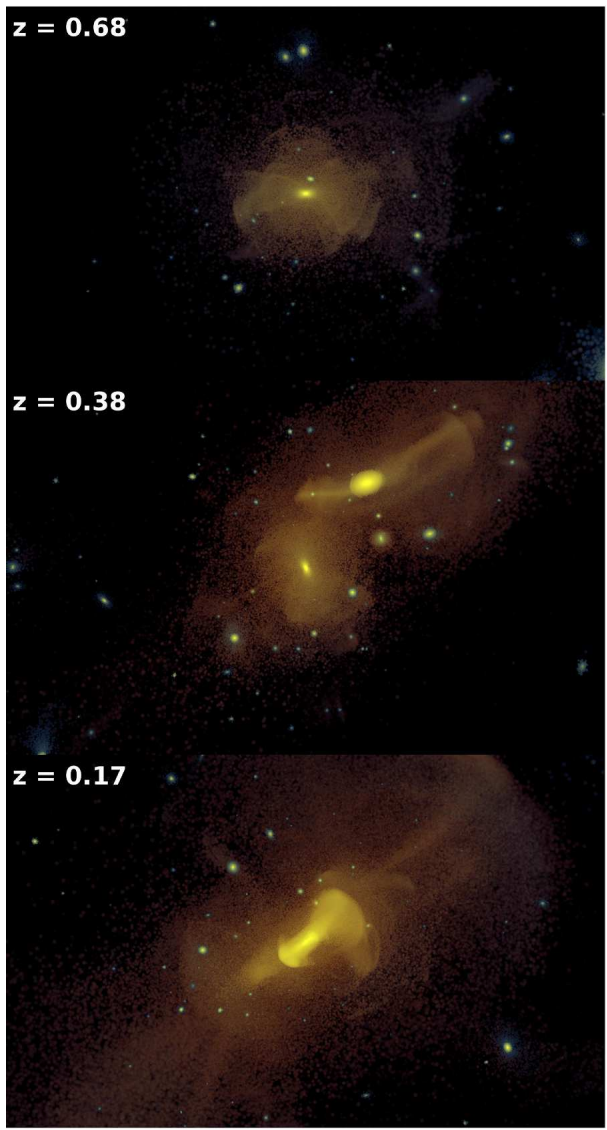}

\caption{The evolution of {\em the stellar halo} in the Aq-F-2 simulation
\citep{Cooper10}. These are still frames from the movie available in the
electronic edition of the Journal, and show stages in the formation of
the shell system at redshifts $z=0.68, 0.38$ and $0.17$, corresponding
to lookback times of 6, 4 and 2~Gyr respectively. The final state is
shown in Fig.~\ref{fig:compare}. Brightness corresponds to projected
stellar mass surface density. Colors, ranging from dark blue to
yellow, correspond to velocity dispersion (from
$\sim50$--$315\,\mathrm{km\,s^{-1}}$). All images are centered on the
main halo and the field of view in each image is approximately 256~kpc
by 160~kpc (comoving). Only accreted stars are shown; the
majority of the stars in the central galaxy do not appear in
these images -- see text and the caption of Fig.~\ref{fig:compare} for
further details.}

\label{fig:movie}
\end{figure}

Deep observations of nearby galaxies have uncovered a wealth of faint
circumgalactic streams, shells and other structures
\citep[e.g.][]{McConnachie09, MD10}. Such features are a natural
occurrence in the cold dark matter (CDM) cosmogony, in which the dark
haloes hosting massive galaxies continually accrete and disrupt their
smaller companions. \citet[][hereafter C10]{Cooper10} have carried out
ultra-high resolution simulations of this process using six N-body
models of Milky Way-mass dark matter haloes from the Aquarius project
\citep{Springel08}. Using the {\sc galform} semi-analytic model of
galaxy formation to calculate the epoch and location of star formation
in the simulation, C10 tagged dark matter particles in appropriate
regions of phase-space to follow the dynamical evolution of stars
stripped from the progenitors of these haloes. In this way, they were
able to model the build-up of galactic stellar haloes through the
tidal disruption of satellite galaxies.

In this paper, we present a movie\footnote{See the online edition of
the journal and \url{http://www.virgo.dur.ac.uk/shell-galaxies}} from
one of the six simulations of C10 (Aq-F-2). The movie is a compelling
illustration of the complexity and dynamism of structure formation in
CDM\@. Because of this complexity, full cosmological modeling is
essential -- a conclusion emphasized by our movie.

The stellar halo of Aq-F-2 contains an extensive system of interleaved
`shells'. Examples of stellar haloes with this distinctive morphology have been
known for decades \citep[e.g.][]{Arp66,Malin83} and their fainter analogs seem
to be common in the local universe \citep[e.g.][]{MD10,Tal09}. In this paper we
present a new deep panoramic image of the diffuse light around one such galaxy,
NGC~7600, showing its well-known shell system and previously undetected faint
features. Even though our ab initio simulation was not constrained to reproduce
NGC~7600 in any way, the observations and the simulation are strikingly
similar, suggesting that such systems arise naturally in the CDM model. 

The movie shows how the simulated system is
  formed in a major ($3:1$) merger between a $\sim 10^{12}\, \mathrm{M_{\sun}}$
  halo and its brightest satellite at $z\sim0.4$. At $z=0$, our
  semi-analytic model predicts that this halo hosts an ellipsoidal
  galaxy (bulge-to-total mass ratio $B/T=0.85$) of total stellar mass
  $1.3\times10^{10}\,\mathrm{M_{\odot}}$.

\section{The Movie}
\label{sec:movie}

Three still frames from our movie are shown in Fig 1. The {\sc
galform} model follows the formation of the entire galaxy (disk,
bulge, halo; see C10 for details), but the movie
shows only the evolution of the stars formed in (and stripped out of) all
progenitors of the final galaxy \textit{except for} the main
progenitor. Stars formed {\em in situ} -- 
those that would make up all of the disk and part of the bulge --
are {\em not} included in our
particle-tagging procedure and are \textit{not} shown.

\begin{figure*}
\center
\includegraphics[clip=True,width=17cm]{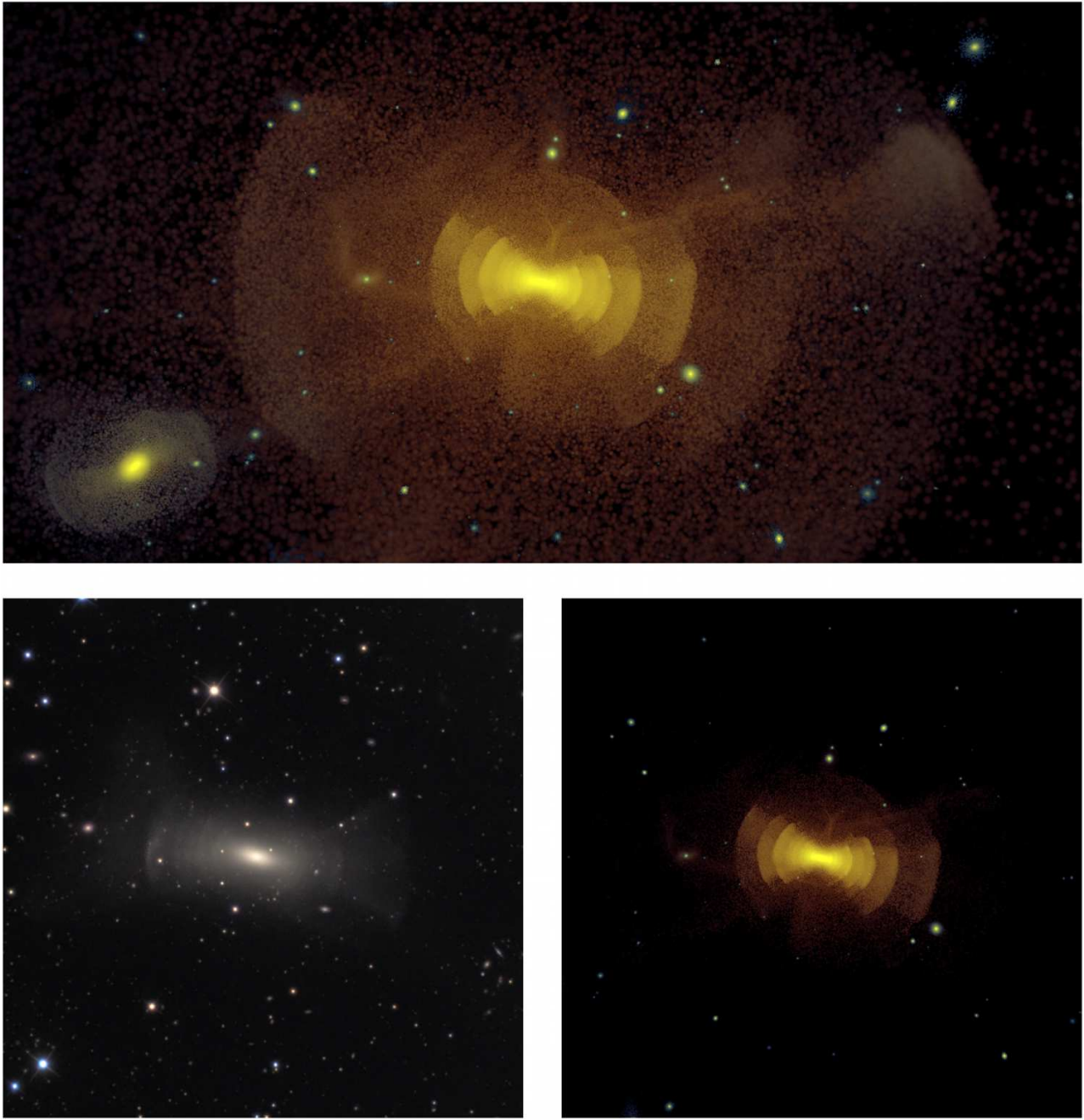}

\caption{{\em Top}: final ($z=0$) frame of the movie. The brightness scale
(surface mass density) was chosen to show the full extent of the structure in
the simulation. Color (from blue to yellow) corresponds to velocity dispersion.
Only {\em accreted} stars are shown so the central concentration of
light corresponds to the small fraction of bulge stars that were stripped from
satellites. {\em Bottom left:} a deep image of NGC~7600 obtained with the
Rancho del Sol 0.5-meter telescope. North is up, East is left. The original
image (see Fig. 3) has been cropped. The total field of view is
$12\arcmin\times 12\arcmin$ ($\sim175\times 175$~kpc). {\em Bottom right:} the
same field of view in Aq-F-2, with a brightness scale chosen to match the
extent of the visible structure around NGC~7600.}

\label{fig:compare}
\end{figure*}

\begin{figure*} 
\begin{center} 
\includegraphics[clip=True,width=17cm]{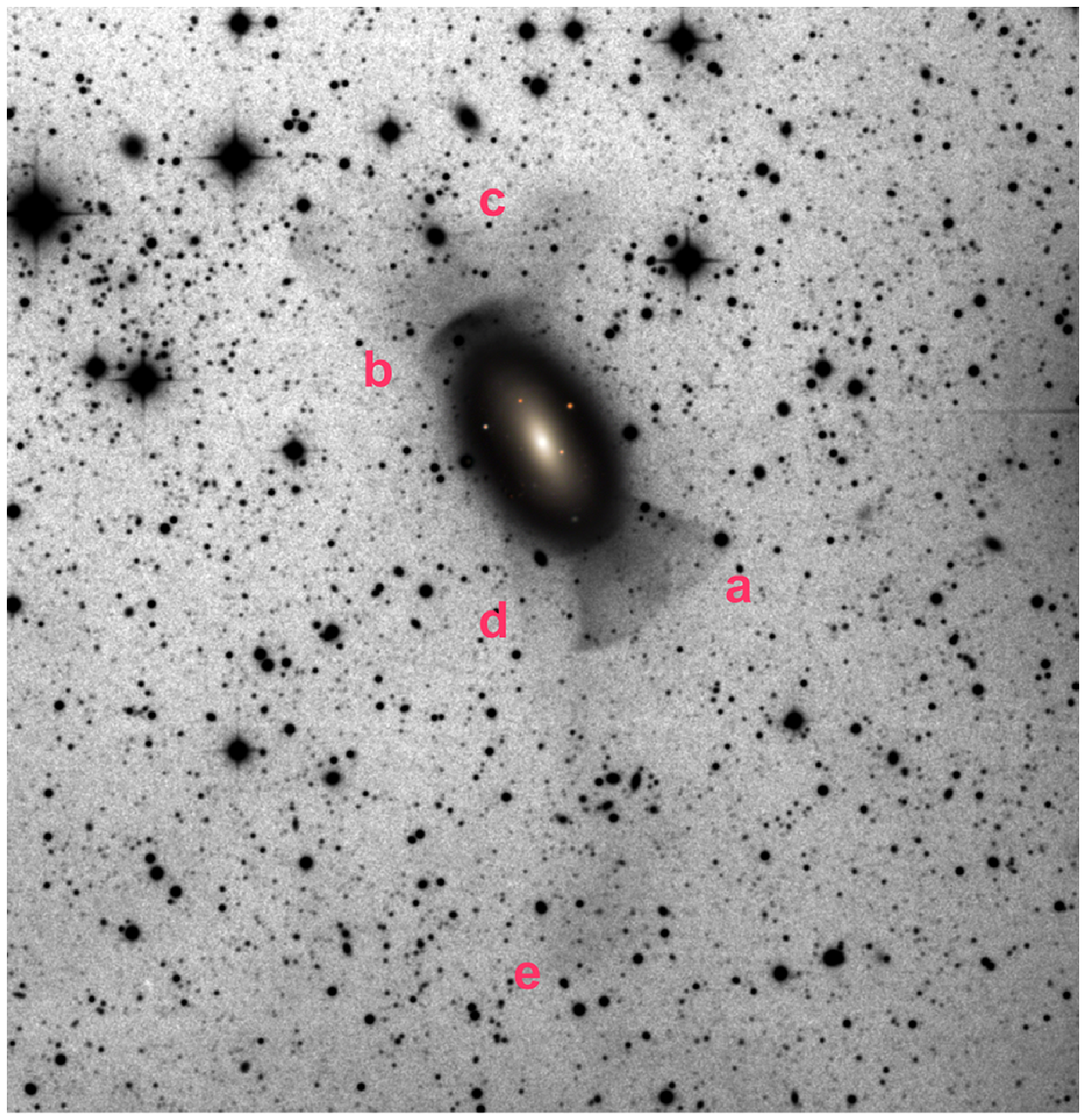}
\end{center}
\caption{A super-stretched, wide field view of the stellar debris around
NGC~7600. The field of view is $18.5\arcmin\times18.9\arcmin$
($268\,\mathrm{kpc}\times 274\,\mathrm{kpc}$). The major axis of the galaxy and
its shells lies along the East-West direction (top left to bottom right). A
color image of the central region of the galaxy is superimposed on the
saturated portion of the image. Labels {\it a} and {\it b} mark the shells
visible in Fig.~2. Other fragments of debris described in the text are labeled
{\it c}--{\it e}. The label {\it e} is 140~kpc in projection from the center of
the galaxy.} \label{fig:widefield} \end{figure*}

The movie begins with the close encounter of two dark matter haloes at
$z=4$. These haloes merge after $\sim2.5$~Gyr ($z=2.7$). Both are
surrounded by a number of dwarf galaxies and the debris of earlier
mergers. The halo entering the picture from below is the more
massive. As the haloes coalesce, their cores oscillate radially about
the center of the potential, creating a series of compact shells
\citep[density caustics at the apocenters of approximately radial
  stellar orbits;][]{Quinn84}. This merger establishes the main halo,
on which the movie is centered for the remainder of the simulation. The
shells propagate rapidly outwards and by $z=2$ have phase-mixed into a
diffuse bow-tie-shaped cloud.

In the next phase of the movie ($z=2-0.5$, spanning $\sim5$~Gyr -- see
the topmost panel of Fig.~1) the halo is bombarded by a number of
smaller satellites on high-angular-momentum orbits. Tidal forces
disrupt many of these satellites, leaving behind streams of debris
that crisscross the halo. As with the shells seen earlier, the stars
in these streams pile up at the apocenters of their orbits. Where the
stream progenitor crosses the center of the halo perpendicular to the
line of sight, it appears as an `umbrella': a broad arc at the end of
a thin stream \citep[e.g.][]{MD10}. These features are rapidly erased
by phase-mixing, perturbations to the potential and the decay of
satellite orbits through dynamical friction.

The final and most spectacular stage of the movie begins at $z=0.4$. As shown
in the central panel of Fig. 1, a bright satellite appears in the upper right
of the frame.

The satellite (whose stellar mass of
$\sim7\times10^{6}\,\mathrm{M_{\odot}}$ at $z=0.4$ is only $\sim0.1$\% of the
stellar mass of the main galaxy at that time, although its dark matter halo is
$1/3$ of the main dark matter halo) seems much brighter than the central galaxy
because only stars {\em accreted} by the central galaxy are shown -- {\sc
galform} predicts that the majority of its stars form in situ, and these are
not tagged by dark matter particles in our model (see C10). The satellite
galaxy brings with it its own extensive stellar halo and set of companions, one
of which is already being disrupted into a wide stream when its host arrives in
the main halo. The bright satellite makes two pericentric passages as the
angular momentum is drained from its orbit. These early passages strip the
satellite of its stellar halo and its companions.  Meanwhile, the stellar halo
of the main galaxy is distorted and mixed by the new arrival, destroying
pre-existing tidal features.

What happens next ($z\sim0.27$, lowest panel of Fig 1.) is crucial for
understanding NGC~7600. The bright satellite is now on an
approximately radial orbit and will pass through the center of the
potential at subsequent pericenters. Material is stripped from the
satellite on each of these passages, just like the stellar
streams. Here, however, liberated stars fan out over a cone of radial
orbits. This is the same phenomenon that created the shells at
$z=2.3$, although here the orbital energy of the satellite is much
greater and the outer shells are much more extensive. As described by
\citet{Quinn84}, stars liberated on a given passage have a range of
binding energies with respect to their host (determined by the
structure of the progenitor and the interaction of the two
potentials). Stars with low binding energies have longer orbital
periods and turn around at larger radii than more tightly-bound
stars. Thus, the outermost shell forms first, from the loosely-bound
stars that lead the satellite along its orbit. All shells move outward
with time as stars with ever-longer periods reach apocentre. At any
given instant, the edges of the shells are composed of stars that have
executed an integer number of orbits \citep[see
e.g.][]{Merrifield98}. They are interleaved in radius on either side
of the center.

Finally, note that the companion that was already disrupting around
the shell progenitor when it entered the main halo embarks on an
independent orbit after the first pericentric passage. This satellite
is continually stripped of its stars, which form a number of umbrellas
with very similar radius and curvature to the shells. At the last
apocentre before its complete disruption ($z\sim0.09$), the debris
stream of this satellite lies perpendicular to the shell axis. The
movie rotates around the halo to show the flattened conical geometry
of the shell system and its dramatically different appearance from
different viewing angles. 

\section{The nearby galaxy NGC 7600}
\label{sec:obs}

NGC~7600 ($D=50\,\mathrm{Mpc}$; $M_{B}=-20.27\pm0.52$) is a nearby elliptical
galaxy \citep[classified visually as S0, but without a disk;][]{Dressler83}.
Its stellar mass, obtained from the method of \citet{Zibetti09}, is
$(6.35\pm1.19)\times10^{10}\,\mathrm{M_{\odot}}$, similar to the stellar mass
of the Milky Way. Deep photographic plates processed by \citet{Malin80}
revealed an interleaved system of sharp-edged shells, which are roughly aligned
with the major axis of the galaxy \citep[][]{Schweizer88}. These shells can be
seen in the lower left panel of Fig.~\ref{fig:compare}, where we present a new
deep image of NGC~7600 obtained using the techniques described in \cite{MD10}.
Our new image also reveals a number of fainter features not previously seen.
Appendix A contains full details of these observations.

The complex field of tidal debris in the outer halo of NGC~7600 is
better seen in the super-stretched, wide field image in
Fig.~\ref{fig:widefield}. In this image the inner shell structure
visible in Fig.~\ref{fig:compare} is saturated, and only two external
shells can be discerned, one on either side of the galaxy. The bright
shell West of the galaxy, labeled {\it a} in Fig.~\ref{fig:widefield},
corresponds to the outermost feature visible in
Fig.~\ref{fig:compare}.  Our deep image reveals two further giant
cones of material West of the galaxy, which are not as clearly aligned
with the major axis. Fragments of these structures were also reported
by \citet{Turnbull99}.

Fig.~\ref{fig:widefield} reveals several features that have not been reported
so far. A diffuse system of debris clouds is visible to the Northeast (labeled
{\it c} in Fig.~\ref{fig:widefield}). It seems to consist of three components,
extending up to 110~kpc from the center of the galaxy. A narrow `spike' emerges
from the diffuse halo (labeled {\it d}). Finally, we detect a large diffuse
stellar cloud (labeled {\it e}) $\sim9.7\arcmin$ from the galaxy, corresponding
to a projected distance of 140~kpc. An extremely faint narrow feature seems to
connect this cloud to the main body of NGC~7600. Together, this diversity of
tidal features suggests that NGC~7600 has undergone an active and complex
merging history in the recent past.

There is an extremely close correspondence between NGC~7600 and the
diffuse structures seen in the Aq-F-2 simulation. Not only are
the morphological similarities striking, but the inner shells in the
simulation have comparable surface brightness to those of NGC~7600,
$\sim27\,\mathrm{mag\,arcsecond^{-2}}$, according to the population
synthesis model used by C10. This is particularly remarkable since the
model galaxy was in no way constrained to resemble NGC~7600.

\section{Discussion}
\label{sec:discuss}

The evolution of Aq-F-2 depicted in the movie highlights several
interesting aspects of the formation of systems analogous to NGC~7600
in a CDM cosmogony.

The first remarkable feature is the complexity of the shell system
itself. The merger origin of circumgalactic shells was first described
in detail by \citet{Quinn84}, who concluded that the shell progenitor
must be a cold stellar system on a radial orbit, as is approximately
the case for the shell progenitor in Aq-F-2. (Few orbits are perfectly
radial to begin with, and, in fact, our shell progenitor passes
pericenter twice before its orbit is aligned and shells are produced.)

The earliest models of shell galaxies were based on simplifying
assumptions and restricted N-body simulations \citep[e.g.][]{Dupraz86}. In the
simplest case, a single instantaneous disruption event liberates all the stars
from the progenitor. The outermost shell is formed first, and successive shells
are created by phase-wrapping. In our simulation, however, the core of the
shell progenitor survives through several pericenters. In the first few
passages, the `new' shells appearing are, in fact, the \textit{first} shells of
\textit{separate} shell systems -- one per pericentric passage of the core.
Only later do phase-wraps of these multiple shells systems begin to appear.

Indeed, the system is even more complex. In the self-consistent
potential of our simulation, energy is exchanged between the shell
progenitor, the main halo and the liberated stars. \citet{Heisler90}
pointed out that interactions between the shell progenitor and its
tidal stream (the `stalk' intermittently visible along the shell axis)
leads to the decay of the satellite's orbit independently of dynamical
friction against the host (both processes take place here). From our
simulation it is clear that the orbit of the satellite decays
significantly before its final disruption.

The evolution in the orbit of the shell progenitor gradually changes
the alignment of new shells throughout the merger. NGC~7600 is highly
elliptical and well-aligned with its innermost shells
\citep{Turnbull99}. The outer shells visible in
Fig.~\ref{fig:widefield} show tentative evidence for misalignment. In
the simulation, the innermost shells (the last to be created) are also
aligned with the major axis of the central bulge. The accreted stars
in the final system have large velocity dispersions
($\sigma>100\mathrm{km\,s^{-1}}$), and exhibit only weak bulk rotation
about the halo center ($v_{rot}\sim10\,\mathrm{km\,s^{-1}}$).

Over the course of the simulation, the morphology of the stellar halo
goes through several short-lived transformations. Many prominent
features are easily disrupted when the central potential is strongly
perturbed. Even the dominant shell system in Aq-F-2 is evolving
rapidly. The first shells created in the merger have already
phase-mixed away by $z=0$. 

Finally, the shell progenitor in Aq-F-2 is a relatively massive dwarf
satellite which enters the main halo along with a number of even
smaller galaxies. This `correlated infall' is characteristic of the
CDM model \citep[e.g.][]{Libeskind05,Li08}. Not only are individual
bright satellites likely to have several luminous companions, but
\textit{most} of these groups, large and small, are accreted along
filaments of the cosmic web. This correlates the directions from which
infalling satellites are accreted with the shape of the dark matter
halo \citep{Lovell11,Vera11}. Some of the satellites of the
shell progenitor are already suffering tidal disruption before the
system is accreted into the main halo. The brightest of these creates
a perpendicular umbrella feature resembling a shell.

\section{Conclusion}
\label{sec:conclusions}

We have presented a new deep image of NGC~7600 revealing new tidal
features in its outer regions. The distribution of stellar debris in
this image is strikingly similar to that of the Aq-F-2 simulation
shown in the movie that accompanies this paper. The simulation
suggests a likely scenario for the formation of the observed shell
system in NGC~7600 and gives a deeper insight into the violent nature
of galaxy assembly in CDM\@.

Our simulation confirms the established hypothesis that shells are created by
the disruption of a satellite system on a radial orbit (here the
satellite is spheroidal, but disk-like progenitors are also possible).
However, in our simulation, the evolution of the orbit of the main shell
progenitor is complex, tidal structures are transient and interact with one
another, and shell-like features are contributed by more than one progenitor.
(Even many of the satellite galaxies around Aq-F-2 themselves show
miniature shell systems.) That we find one such example among six randomly
selected $\sim10^{12}\,\mathrm{M_{\odot}}$ haloes suggests that such mergers
and complex shells are a natural expectation of the CDM model. Deeper
observations of nearby galaxies should reveal many more cases.

Although the prevalence of shell systems and their dependence on galaxy
properties remains to be quantified, both in the real universe and in
simulations, it is clear that these features offer the prospect of a novel test
of the CDM model on galactic scales. These scales are the most sensitive to
fundamental properties of the dark matter, for example to whether the dark
matter is cold or warm. Hierarchical formation -- the hallmark of cold dark
matter -- must be accompanied by the formation of shells, a process that is
testable with deep, wide-field observations of nearby galaxies.

\acknowledgments

DMD thanks the Max-Planck Institute for Astrophysics for hospitality
during the preparation of this work. CSF acknowledges a Royal Society
Wolfson Research Merit Award and ERC Advanced Investigator grant
267291 COSMIWAY\@. SMC acknowledges a Leverhulme Fellowship. The Dark
Cosmology Centre is funded by the Danish National Research
Foundation. Calculations for this paper were performed on the ICC
Cosmology Machine, part of the DiRAC Facility, the Leibniz
Rechnungszentrum (Garching) and LOFAR/STELLA (Groningen). This work
was supported in part by an STFC rolling grant to the ICC.

\appendix

\section{Observations and data reduction}
\label{sec:Data}

We have obtained new deep optical images of NGC~7600 with the 0.508~m
Ritchey-Chr\'etien telescope of the Rancho del Sol Observatory in
Camino, California. We used an Apogee Alta u9000 CCD camera (pixel
size 12 micron), providing a field of view of
$29.3\arcmin\times29.3\arcmin$ at a plate scale of
$0.58\arcsec\,\mathrm{pixel}^{-1}$. Our image set consists of 34
individual exposures of 20 minutes each with a clear luminance filter
($3500<\lambda<8500$). These dark-sky images were obtained between
September 2nd and September 9th, 2010.  They were reduced using
standard procedures for bias correction and flatfield, as explained in
\cite{MD09}. The final image was obtained by summing all the luminance
CCD exposures, with a total accumulated exposure time of 680~minutes
(11.33~hours).

We performed a careful subtraction of the background in the final
reduced image. On large scales, the background is heavily affected by
the scattered light of bright sources and other artifacts (possibly
due to internal reflections in the camera). For this reason, we
optimize the subtraction within $\sim8\,\arcmin$ from the galaxy
center. We select 36 boxes around the galaxy, chosen not to include
bright stars or be contaminated by the galaxy halo and the faint
diffuse features. A first degree 2-dimensional polynomial is fit to
the pixels in these regions using a very conservative sigma-clipping,
and subtracted from the image to remove background gradients. We
further subtract the residual pedestal given by the median counts of
pixels in the selected regions.

An obvious disadvantage of luminance images is the lack of a
photometric standard system for direct calibration of the flux. For
this purpose we use images of the same sky area from the Sloan Digital
Sky Survey \citep[SDSS DR8,][]{Aihara11} to calibrate our luminance
image in the SDSS photometric system (specifically the $g$, $r$ and
$i$ bands that cover the spectral range of the luminance filter). To
reduce the uncertainty due to the unknown color terms, we calibrate
our image on the brightest regions of NGC~7600 (within
$\sim1\,\arcmin$) and assume that the faint features have similar
colors to the main galaxy body
\citep[as suggested by][]{Turnbull99}. From the noise properties of
the background we derive a $3\sigma$ threshold for feature detections
over $2\arcsec$ apertures of 28.1, 27.3 and 26.9
$\mathrm{mag}\,\mathrm{arcsec}^{-2}$ in $g$, $r$ and $i$ bands,
respectively. The $1\sigma$ accuracy of the background subtraction is
estimated from the box-to-box variance of the average residual surface
brightness in 29.7, 28.9 and 28.5 $\mathrm{mag}\,\mathrm{arcsec}^{-2}$
in $g$, $r$ and $i$ bands, respectively.

We derive a mass-to-light ratio from the $H$-band luminosity and ($g-i$,$i-H$)
colors of NGC~7600, using the lookup tables of \citet{Zibetti09}, and estimate
a stellar mass of $6.35\times10^{10}\pm1.91\,\mathrm{M_{\odot}}$. $H$ is taken
from 2MASS, $g$ and $i$ from SDSS\@. All photometry is integrated within the
elliptical isophote corresponding to $25 \mathrm{mag\,arcsec^{-2}}$ in the SDSS
$r$ band ($150\min$ semi-major axis, $b/a=0.39$). We assume a distance modulus
of $33.18$.  Photometric quantities (corrected for foreground galactic
extinction) are: $m_g=12.12$, $m_r=11.44$, $m_i=11.03$, $m_H = 8.79$ (Vega).
The H-band luminosity is $1.27\times10^{11}\,\mathrm{L_{\odot}}$.





\end{document}